\documentclass[10pt]{article}
\usepackage{graphicx}
\def\ls{\mathrel{\raise1.16pt\hbox{$<$}\kern-7.0pt 
\lower3.06pt\hbox{{$\scriptstyle \sim$}}}}

\def\gsim{\;\rlap{\lower 2.5pt
 \hbox{$\sim$}}\raise 1.5pt\hbox{$>$}\;}
\def\lsim{\;\rlap{\lower 2.5pt
   \hbox{$\sim$}}\raise 1.5pt\hbox{$<$}\;}

\newcommand{\beq}{\begin{equation}}
\newcommand{\eeq}{\end{equation}}
\def\myputfigure#1#2#3#4#5
{\hskip0.03\textwidth\vskip#5pt
\makebox[0pt]{\hskip#2in
\includegraphics[width=#3\textwidth]{#1}}\vskip#4pt\hfill}


\begin{document}

\hrule
\vskip 0.0in
\Large{Significant primordial star formation at redshifts $z\approx 3-4$.}

\normalsize \author{Raul Jimenez$^1$, Zoltan Haiman$^2$}

Raul Jimenez$^1$, Zoltan Haiman$^2$

\noindent
$^1$Department of Physics and Astronomy, University of Pennsylvania, 209 South 33rd Street, Philadelphia, PA 19104, USA; raulj@physics.upenn.edu\\
$^2$Department of Astronomy, Columbia University, 550 West 120th Street, New York, NY 10027, USA; zoltan@astro.columbia.edu \vskip 0.2in \hrule \vskip 0.2in

\noindent

{\bf Four recent observational results have challenged our
understanding of high--redshift galaxies, as they require the presence of far more 
ultraviolet photons than should be emitted by normal stellar populations. First, there is significant ultraviolet
emission from Lyman Break Galaxies (LBGs) at wavelenghts shorter than
$912$\AA\cite{setal01}. Second, there is strong Lyman $\alpha$ emission from
extended ``blobs'' with little or no associated apparent ionizing
continuum\cite{matsudaetal2004}. Third, there is a population of galaxies with
unusually strong Lyman $\alpha$ emission lines\cite{malhotra02}.  And fourth, there is a strong HeII (1640 \AA) emission line in a composite of
LBGs\cite{shapley03}. The proposed explanations for the first three observations are internally inconsistent, and the fourth puzzle has remained hitherto unexplained. Here we show that all four problems are resolved simultaneously if $10-30$ percent of the stars in many galaxies at $z\approx 3-4$ are mainly primordial -- unenriched by elements heavier than helium ('metals').  Most models of hierarchical galaxy
formation assume efficient intra--galactic metal mixing, and therefore
do not predict metal-free star formation at redshifts significantly below $z \sim
5$
\cite{madauetal01,Scannapiecoetal03,Normanetal04,Yoshidaetal04,BrommLoeb05}.
Our results imply that micro--mixing of metals within galaxies is
inefficient on a $\sim$ Gyr time--scale, a conclusion that can be
verified with higher resolution simulations, and future observations
of the HeII emission line.}

The continuum emission from LBGs can be explained by adding the
continuous formation of massive stars to the standard initial mass
function (IMF).  In the absence of such a bias in the IMF, the
continuum flux emerging just below $< 912$ \AA\ will be dominated by
low--mass stars and will be significantly fainter than observed.
The other three puzzles require a substantial increase
in the total number of energetic ionizing photons, and we find they
cannot be resolved by a bias of the IMF alone. To increase the number of 
ultraviolet photons further the
temperature in the atmospheres of the stars must be increased. This is the case for
metal-free (often called population-III) stars, which have
significantly reduced opacity owing to the lack of CNO elements \cite{TS2000,KBL2001}.

Lyman-continuum flux has been detected \cite{setal01} in the composite spectrum of 29 Lyman Break
Galaxies (LBGs) at a mean redshift of $z\approx 3.4$.  
Although the authors of ref. 1 warn that sky background subtraction at the 
low observed flux level is a concern, when taken at face value, 
their  result
implies that a surprisingly large fraction of ionizing photons can
typically escape from these galaxies. These authors found $f_{\rm
esc}\gsim 0.5$, where $f_{\rm esc}$ is defined as the ratio of
$900$\AA\ photons that escape the galaxy and the $1500$\AA\ photons
that escape (the wavelengths are quoted in the rest--frame of the
LBGs).  Using more detailed fits of the observed spectrum,
\cite{huietal02} and \cite{haehnelt01} found  that the spectrum
appears surprisingly blue, even if {\em no} additional Lyman continuum
absorption is assumed to take place in the galaxy (beyond the dust
absorption that also affects the spectrum at $1500$\AA).

We compare the stacked composite spectrum of LBGs in \cite{setal01} 
with model galactic spectra, including metal--free
stars. The intrinsic emission of the galaxies is modeled as the sum of
``normal'' star--formation (stellar populations with metallicity ratio $>
10^{-3} Z/{\rm Z_{\odot}}$ \cite{Brommetal01,Schneideretal02}), 
and metal--free, massive pop--III
star--formation.  For the ``normal'' component of the intrinsic
stellar emission, we use the spectral synthesis models of
\cite{starburst99} for a Salpeter IMF (with a slope $\alpha=2.35$)
between $1-100$M$_\odot$ and a metallicity of $Z=0.4$Z$_\odot$, and
compute the spectrum for continuous star formation for $10^8$ yr,
which is the age for which a steady-state has been reached.  For the
`metal--free' component, we use the models of \cite{schaerer03},
and adopt the same Salpeter IMF, but limited to massive stars in the
$100-500$M$_\odot$ range \cite{AbelBryanNorman02,BrommCoppiLarson02}, and with $Z=0$.  Note that massive stars
with $>100$M$_\odot$ have similar spectra, and our results should not
depend significantly on the distribution of pop--III stellar masses above
this threshold value. To simulate the observed, processed composite spectrum, we modify the
emitted stellar template spectrum by including the effects of dust
absorption, as well as the opacity of the intervening intergalactic medium (see Supplementary information). 

As the top panel in Fig.~\ref{fig:fesc} shows, in the $880$\AA$\leq
\lambda\leq 910$\AA\ range, the mean observed flux is quite high,
$F_\nu/F_{1500}=0.064\pm 0.013$; while the normal--only star formation
model predicts a much lower value of $F_\nu/F_{1500}=0.023\pm
0.01$. Taken at face value, the data reveals an ionizing flux about three times higher than this model predicts, even if we do not include any
continuum absorption by neutral gas in the galaxy.  The data and the
model do not agree at the $>3\sigma$ level.  The model with pop--III
star--formation with $f_{\rm popIII}=0.5$ almost generates the observed ionizing flux in the $880$\AA$\leq \lambda\leq 910$\AA\
range; producing a mean $F_\nu/F_{1500}=0.048\pm 0.02$.

Our conclusions regarding the LBGs apply to the bluest
sub-sample. This is further supported by the analysis of the CIV
absorption line associated with the stellar winds in massive stars
that is observed in the spectra of LBGs. According to our prediction,
the bluest LBGs should have weaker CIV absorption than the reddest
LBGs, because pop-III stars are metal free and therefore contribute
significantly fewer metals. 
This trend is indeed what is observed (see table 3 and section 5.3 in
\cite{shapley03}). Detailed inspection of the numbers in table 3 of
\cite{shapley03} shows that the weaker strength of the CIV line
requires an abundance of $ \sim 50$\% pop-III in the bluest LBGs. 
We note that the CIV profile in LBGs is a complex superposition of
interstellar absorption, stellar-wind emission and absorption, and
possibly nebular emission.  Our models predict that the trend of
decreasing CIV strengh in the bluest galaxies is present in the narrow
interstellar absorption  and any broad stellar component (which measures the total cumulative C output of all stars), but not necessarily in any of the
other components. 
The fact that metal lines are observed in all
LBGs indicates that LBGs are not composed purely of pop-III stars, but
that there is a mixture in each galaxy of pop-III and normal stars.

There have been recent observations of a population of
$35$ Ly$\alpha$ blobs at $z=3.1$ in narrow and broad band spectra\cite{matsudaetal2004}.  For about one third of the blobs, the
amount of photons provided by a normal stellar population that fits the observed spectrum at
wavelengths longer than $912$ \AA is not capable of producing the
required number of ionizing photons to explain the observed Ly$\alpha$
flux. If we assume case B (two Ly$\alpha$ photons are produced for
every three ionising photons from the stars) then a stellar population
with $Z=10^{-2} {\rm Z_{\odot}}$ would produce at least 35\% less photons than those
necessary to account for the observed Ly$\alpha$ flux. To reproduce the observed flux, about 30\% of the stars
would need to be metal-free.

Figure~\ref{fig:blobs} shows that about $20-30$\% of the
blobs also have continuum colors consistent with the stars producing the ionizing
Ly$\alpha$ photons being metal-free. The fact that the stellar
populations powering the extended emission are metal--poor is consistent with the fact that these sources are
extended, and are presumably more likely to have been caught in the
process of their assembly, when they are less metal--enriched overall
.  They could plausibly be identified, therefore, as the progenitors
of LBGs. The rest-frame 
equivalent width for Ly$\alpha$ ranges from $20$ to $350$ \AA. For models with constant star formation and $Z > 10^{-5} {\rm Z_{\odot}}$ the EW are $70-100$ \AA. About 
one third of the blobs have EW larger than $100$ \AA\, and therefore require metal-free stars. This is in agreement with what is found for the colors, although we note that there is little overlap among
the individiual blue blobs and those with Lya excess. The good fit to the $B-I$ colors,
without including (case-B) reprocessing of ionizing radiation
into Lyman $\alpha$, implies that $f_{\rm esc}$ is of the order of $100$\% because a smaller escape fraction would imply more flux on $B$ and therefore significantly bluer $B-I$ colors than those observed.  The large escape fraction of ionizing radiation would be consistent with the value observed in the composite of LBG spectra.

Strong HeII (1640 \AA\,) emission line from a composite of LBGs has
been observed \cite{shapley03}. They report
equivalent widths for a sample of several hundred LBGs of about $2$
\AA. This line is also very broad, with a FWHM of about $1500$ km s$^{-1}$. Although this cannot
be explained by stars without winds, it has been shown
(e.g. \cite{marigoetal03,meynetetal05}) that, aided by rotation, metal--free stars
can loose significant amounts of mass and drive strong winds, similar
to normal--metalicity stars. In this case, the He--ionizing photons
from the massive popIII stars will be reprocessed in the optically
thick stellar outflows surrounding the star, making the line broad \cite{Kudritzki}. 
As discussed in \cite{shapley03}, 
normal WR stars cannot explain the HeII line observed in
the composite LBG sample, because the requisite number of WR stars
will overproduce the stellar CIV emission. 

However, metal-free massive stars with winds do not have this problem. As shown in \cite{Kudritzki} there is a clear anti-correlation between the strength of the HeII 1640 line and the CIV line (see Fig. 13 in \cite{Kudritzki} for the case $Z=0.01$ and $10^{-4}  Z_{\odot}$). For metal-free stars the equivalent width of the HeII 1640 line for a continuous star forming ($1$
M$_{\odot}$ yr$^{-1}$) model of metal-free stars at age $0.1$ Gyr is $30$ \AA, significantly larger than observed. So for
this sample, the required amount of metal-free stars is only
$10$\%. However, the equivalent width of the HeII 1640 line is very
sensitive to the metallicity.  If we use $Z=10^{-7} {\rm Z_{\odot}}$
(instead of $Z=0$), then the $912$ \AA\, decrement doesn't change, so
our conclusions above about the UV radiation from LBGs and the Ly$\alpha$ blobs
remain unchanged. But the He-ionizing flux, and the He1640 equivalent
width is reduced by about a factor of $6-7$. This implies a population of $\sim 30$\% metal-free stars. 

Recent non--detections of the HeII 1640 line in a composite spectrum
of 17 galaxies at $z=4.5$~\cite{Dawsonetal04}, as well as from an
individual Ly$\alpha$ emitter at $z=6.33$~\cite{Nagaoetal05} have
provided tight upper limits, but still allow for a significant mass
fraction ($\sim50\%$) in Pop III stars.

The Large Area Lyman Alpha survey \cite{malhotra02} reports large
Ly${\alpha}$ equivalent widths for a sample of 150 emitters at $z
\approx 4.5$. The median value of the distribution is $\sim 450$ \AA\,
and 60\% have equivalent widths above $240$ \AA . No continuum
emission was found in 30\% of the emitters. However, the equivalent
widths are spread over a wide range: from 10 to 5000 \AA. This implies
that not a single type of stellar population can account the large
spread of equivalent widths. The first suggestion that this large equivalent widths can be produced by very low metallicity stars is found in \cite{Kudritzki2}. We find
that for metal-free stars and constant star formation, the equivalent
width is $1000$\AA\, for the IMF used in this paper and declines to
$300$\AA\, for a Salpeter IMF \cite{schaerer03}.  We performed a more detailed fit to
the distribution in \cite{malhotra02}. Using models for
metallicities $Z > 10^{-5} {\rm Z_{\odot}}$ we find that bursts of ages
greater than $10^7$ years have equivalent widths as large as 300\AA\,
whereas constant star formation models have about 200\AA. Therefore
"normal" stellar populations can account for as much as 40\% of the
Ly${\alpha}$ emitters, while the rest need to be accounted for by
metal-free stellar populations. 
 
Recently two stars with metallicity $Z < 10^{-5} {\rm Z_{\odot}}$ have been
found in the Milky Way \cite{Christliebetal02}. Although
such low--mass stars ($\lsim 0.8$ M$_{\odot}$) are not included in our
model, they reveal the details of the earliest
stages of chemical enrichment, and suggest an initial overproduction
of Carbon. For stars with $[Fe/H] < -3.5$, 40\% have $[C/Fe] > 1.0$
\cite{BeersChristlieb05} and the trend increases to 100\% for stars
with $[Fe/H] < -5$. This trend can naturally arise in our model  \cite{marigoetal03,meynetetal05}. 

Our results imply that micro--mixing of metals within galaxies is
inefficient on a $\sim$ Gyr time--scale, a conclusion that may be
surprising, as current models of hierarchical galaxy formation
assume efficient intra--galactic metal mixing, and therefore do not
predict metal-free star formation significantly below $z \sim 5$
\cite{madauetal01,Scannapiecoetal03,Normanetal04,Yoshidaetal04,BrommLoeb05}.
However, the presence of significant pockets of metal--free
star--formation explains four puzzling observational results, and the
hypothesis does not violate any other existing observations.  The
presence of metal--free gas pockets can be verified in future high
resolution numerical simulations, and by future observations. 
Our models predict that in some of the LBGs there should be a detectable
continuum shortward of 912\AA\, and a detectable individual HeII
line. Long--integration observations of these individual galaxies in
the future could be used tocharacterize their spectra, and to confirm
the presence of metal-free stellar populations. In particular we predict that there
should be a clear anti-correlation between He1640 and CIV line
strength and this should be a reliable test of the amount of metal-free stars at $z \sim 3-5$.

Supplementary information accompanies the paper on www.nature.com/nature.\\

{\bf Acknowledgements:} 
We thank Y. Matsuda and T. Yamada for providing the
unpublished raw color data that appears in Figure 2,
D. Schaerer for providing pop-III spectra in electronic
form, and A. Shapley for useful discussions. We also thank  T. Beers, for pointing out the C trends in low metallicty stars and suggesting these could be explained by our model. RJ and Z.H. gratefully acknowledge financial support from NSF  and NASA.

\clearpage

\noindent {\bf Figure 1:} {\bf Lyman break galaxies stacked observed spectrum and best fit population synthesis models.} The dotted curve corresponds to continuous
normal star formation for $10^8$ years, and the dashed curve
corresponds to a mix model where 50\% of the stars (by mass) are
massive, metal--free stars in the $100-500~{\rm M_{\odot}}$ range. The
solid curves show the composite spectrum in \cite{setal01}.  All
fluxes are normalized to the flux at the emitted wavelength of
1500\AA.  For reference, in the bottom panel, we show the optical
depths assumed for the dust and the IGM.  In the top panel, the
horizontal dashed line shows the mean observed flux in the interval
$880$\AA$\leq \lambda\leq 910$\AA.  The inference is that the spectrum
requires a significant contribution by metal--free star--formation,
with a mass--fraction $f_{\rm popIII}\approx 0.5$. As discussed in \cite{huietal02}, changing the assumed metallicity of the normal stars does not
significantly increase the predicted ionizing fluxes.  The ultraviolet fluxes
of OB stars can be increased by the presence of stellar winds (not
included in the stellar models we adopted).  Although winds can
increase the HeII--ionizing flux by orders of magnitude, the
corresponding increase for the H--ionizing flux in O stars has been
found to be small \cite{schaerer97}.  The increase can be
more significant for cooler B--stars \cite{najarro96}, but these
stars do not dominate the ionizing photon budget in the continuous
star--formation models, in which O--stars are continuously
replenished.  Furthermore, the hydrostatic stellar models typically
reproduce the properties of Galactic HII regions \cite{starburst99}, 
so that an increase by the required factor of $\sim 2$
would make it more difficult to reconcile the models with these
observations.

\noindent {\bf Figure 2:}  {\bf Colour distributions of the Ly$\alpha$ blobs.} The different lines correspond to
stellar population models with continuous star formation 
at an age of $10^8$ yr for $Z=0.2 {\rm Z_{\odot}}$
(dashed), $Z=0.01 {\rm Z_{\odot}}$ (dotted) (both computed using the models
from \cite{Jimenezetal2004}) and $Z=0$ (solid, assuming an IMF and
mass range $100-500~{\rm M_\odot}$ as in our analysis of LBG spectra).  
We have included the effect of the opacity of the intervening
intergalactic medium.  
The shaded ranges indicate the range of expected colors as various properties of the
metal--free population are varied, such as their age (between
$10^{5}-10^{8}$yrs), the IMF (for the three models given in
ref.~\cite{schaerer03}), and star--formation history (between a burst
and continuous star formation at a constant rate). 
All three colour histograms suggest that about 20\% of the blobs need to be made of purely popIII stars.
Note that no correction was made in the LAB colours for
Lya line emission or for possible continuum contribution
from foreground and background objects.

\clearpage

\begin{figure} 
\includegraphics[width=14cm,height=10cm]{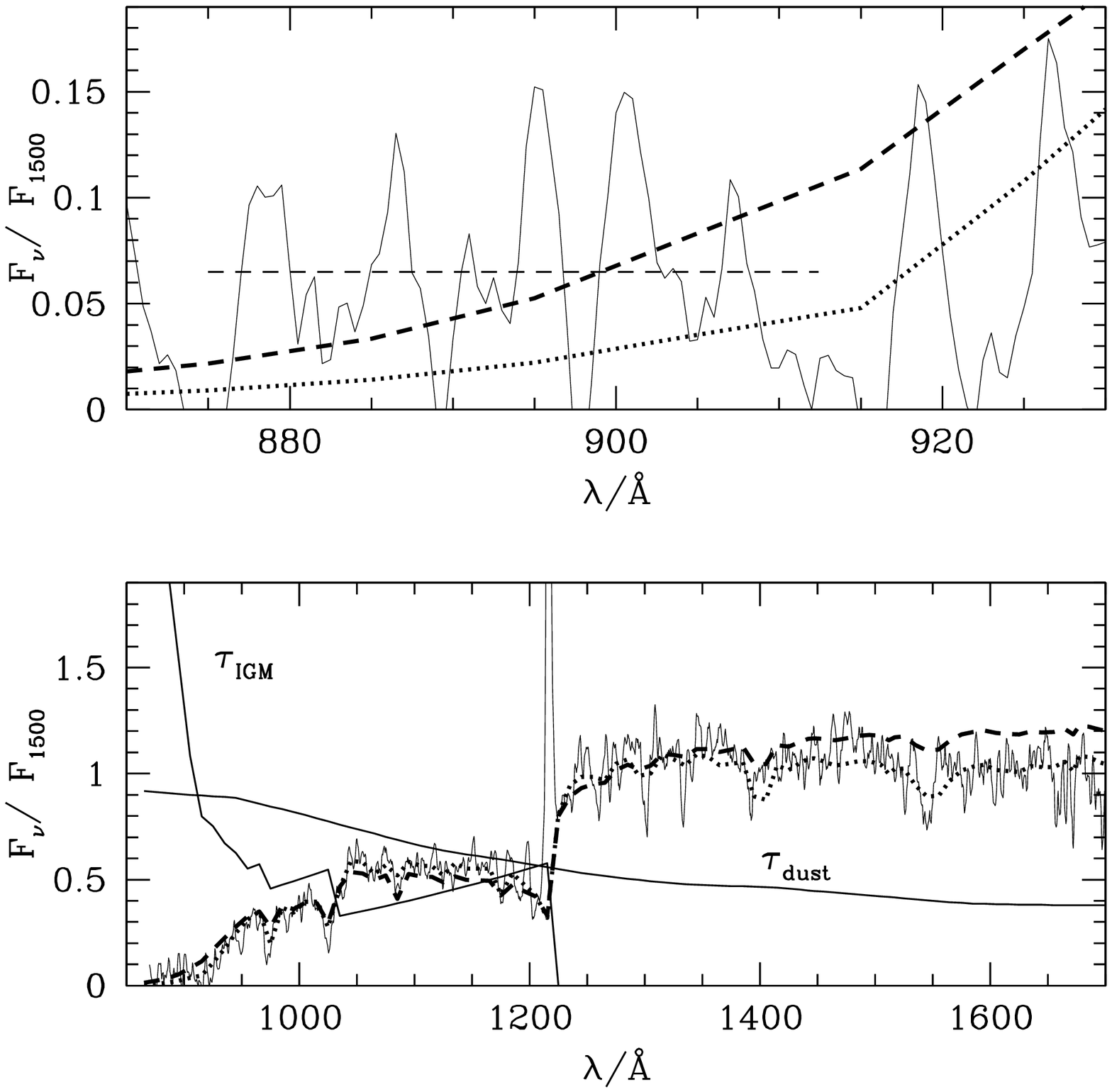}
\caption{\label{fig:fesc}}
\end{figure}

\clearpage

\begin{figure}
\includegraphics[width=10cm,height=14cm]{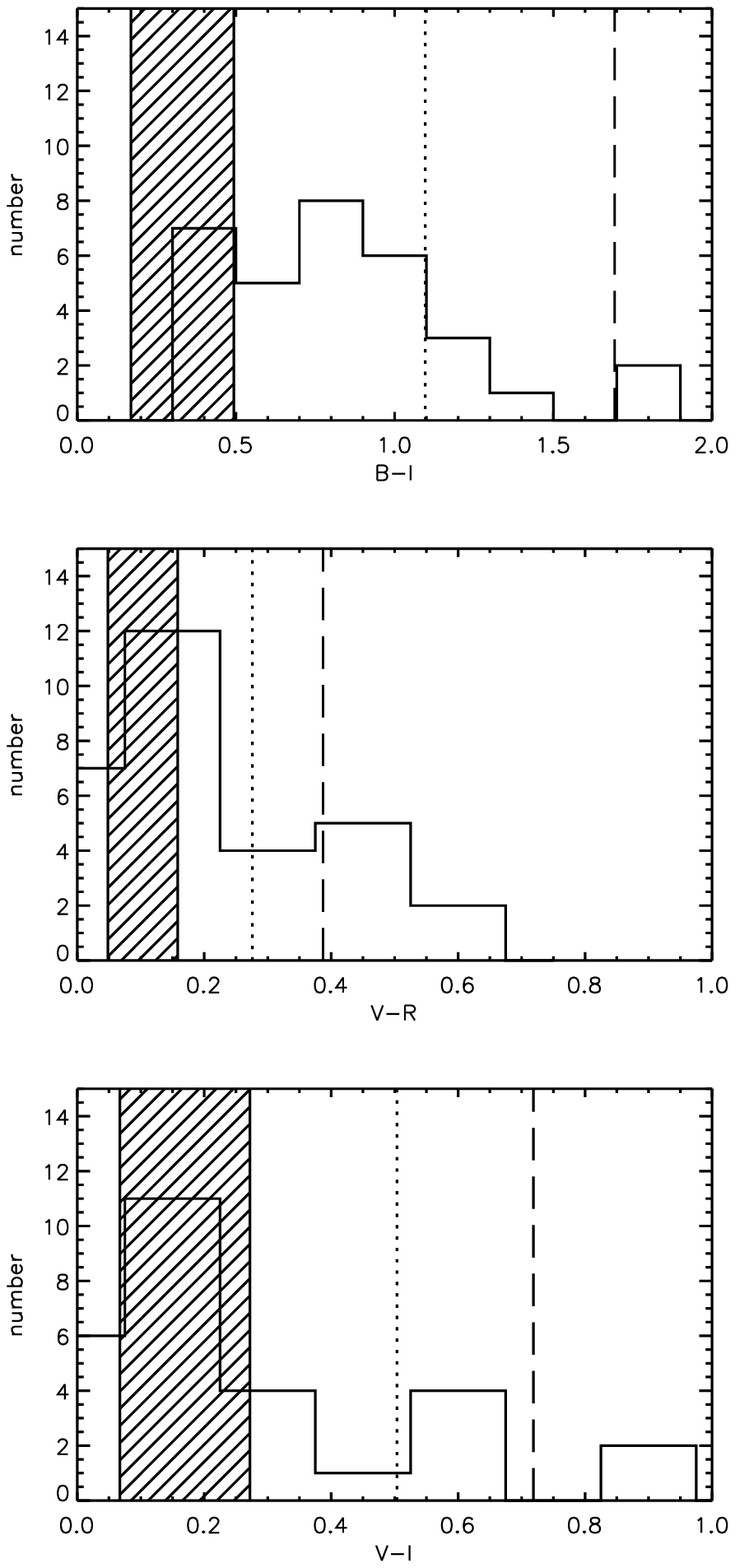}
\caption{
\label{fig:blobs}}

\end{figure}

\end{document}